\documentclass[twocolumn,aps,prl,nofootinbib,showpacs,twoside,notitlepage,superscriptaddress]{revtex4-2}

\usepackage{amsmath, hyperref}
\usepackage{amssymb}
\usepackage{amsfonts}
\usepackage{enumerate}
\usepackage{graphicx}
\usepackage{bm}
\usepackage{hyperref}
\usepackage{graphicx}
\usepackage[dvipsnames]{xcolor}
\usepackage{tikz}
\usetikzlibrary{decorations.pathmorphing}
\usepackage{ulem}

\DeclareMathOperator{\sgn}{sgn}

\newcommand{\ham}{\mathcal{H}}
\newcommand{\diff}{\mathcal{D}}
\newcommand{\shift}{{N^x}}
\newcommand{\lapse}{N}
\newcommand{\erad}{E^x{}}
\newcommand{\ephi}{E^\varphi{}}
\newcommand{\krad}{K_x}
\newcommand{\kang}{K_\varphi}

\def\exterior{E}
\def\interior{I}
\def\uint{\bar u}
\def\vint{\bar v}
\def\Sunit{d\Omega^2}
\def\rext{\tilde{r}}
\def\rint{\bar{r}}

\def\ff{\Gamma}
\def\horizon{\mathcal{Z}}
\def\xu{z}
\def\time{t}
\def\radi{x}
\def\timeI{\overline T}
\def\xuI{\overline Y}

\begin{document}

\title{An effective model for the quantum Schwarzschild black hole}

\author{Asier Alonso-Bardaji}
  \email{asier.alonso@ehu.eus}
	\affiliation{
 	Fisika Saila, Universidad del Pa\'is Vasco/Euskal Herriko Unibertsitatea (UPV/EHU),
	Barrio Sarriena s/n, Leioa, Spain}

\author{David Brizuela}
  \email{david.brizuela@ehu.eus}
	\affiliation{
Fisika Saila, Universidad del Pa\'is Vasco/Euskal Herriko Unibertsitatea (UPV/EHU),
Barrio Sarriena s/n, Leioa, Spain}
	
\author{Ra\"ul Vera}
  \email{raul.vera@ehu.eus}
	\affiliation{
	Fisika Saila, Universidad del Pa\'is Vasco/Euskal Herriko Unibertsitatea (UPV/EHU), Barrio Sarriena s/n, Leioa, Spain}

\begin{abstract}

We present an effective theory to describe the quantization of
spherically symmetric vacuum motivated by loop quantum gravity.
We include anomaly-free holonomy corrections through a canonical transformation and a linear combination of constraints 
of general relativity,
such that the modified constraint algebra closes.
The system is then provided with a fully covariant and unambiguous
geometric description, independent of the gauge choice on the phase space.
The resulting spacetime corresponds to a singularity-free
(black-hole/white-hole) interior and two asymptotically flat exterior regions of equal mass.
The interior region contains a minimal smooth spacelike surface that replaces the Schwarzschild singularity.
We find the global causal structure and the maximal analytical extension.
Both Minkowski and Schwarzschild spacetimes are directly recovered as particular limits of the model.
\end{abstract}

\maketitle

The singularities predicted by general relativity (GR)
are expected to disappear once a complete quantum
description of gravity is achieved. Loop quantum gravity
predicts a quantized spacetime presumably
mending those defects. However, a complete quantum description of the regions close
to a singularity is not at hand and one must consider effective descriptions that implement
the expected corrections.
In particular, the accuracy shown by effective techniques
for homogeneous models \cite{cite-key,Ashtekar:2011ni,Agullo:2016tjh,Ashtekar:2021kfp},
where the initial singularity is replaced by a quantum bounce,
has been the
motivation to extend the so-called holonomy corrections to
spacetimes with less symmetry.

Concerning non-homogeneous models, the most simple scenario is that of a spherically symmetric
black hole. The main approach in the literature has dealt just with its interior
part by using the same techniques as for homogeneous models \cite{Boehmer:2007ket,Chiou:2008eg,Chiou:2008nm,Joe:2014tca,Olmedo:2017lvt}.
Nonetheless, the implementation of the isometry between the homogeneous interior and
Kantowski-Sachs cosmology is only partially satisfactory 
and a comprehensive geodesic analysis is mandatory.
In this respect, 
there are several
proposals \cite{BenAchour:2018khr,Ashtekar:2018cay,Ashtekar:2018lag,Bodendorfer:2019cyv,Bodendorfer:2019nvy,Gambini:2020nsf,Kelly:2020uwj} which, however, present crucial problems that we address in our model.
For instance, the extension to the exterior static region, the asymptotic flatness, the slicing-independence and the confinement of quantum effects to large-curvature regions
are open issues present in most of the models in the literature.
Moreover, none of the mentioned studies addresses explicitly
the covariance of the theory \cite{Bojowald:2015zha,Bojowald:2018xxu,Bojowald:2020dkb,Bojowald:2020unm,Alonso-Bardaji:2020rxb,Bojowald:2021isp}: quantum effects may thus depend on the particular gauge choice
and not yield conclusive physical predictions.

\begin{figure*}[ht!]
  \centering
  \includegraphics[width=\textwidth]{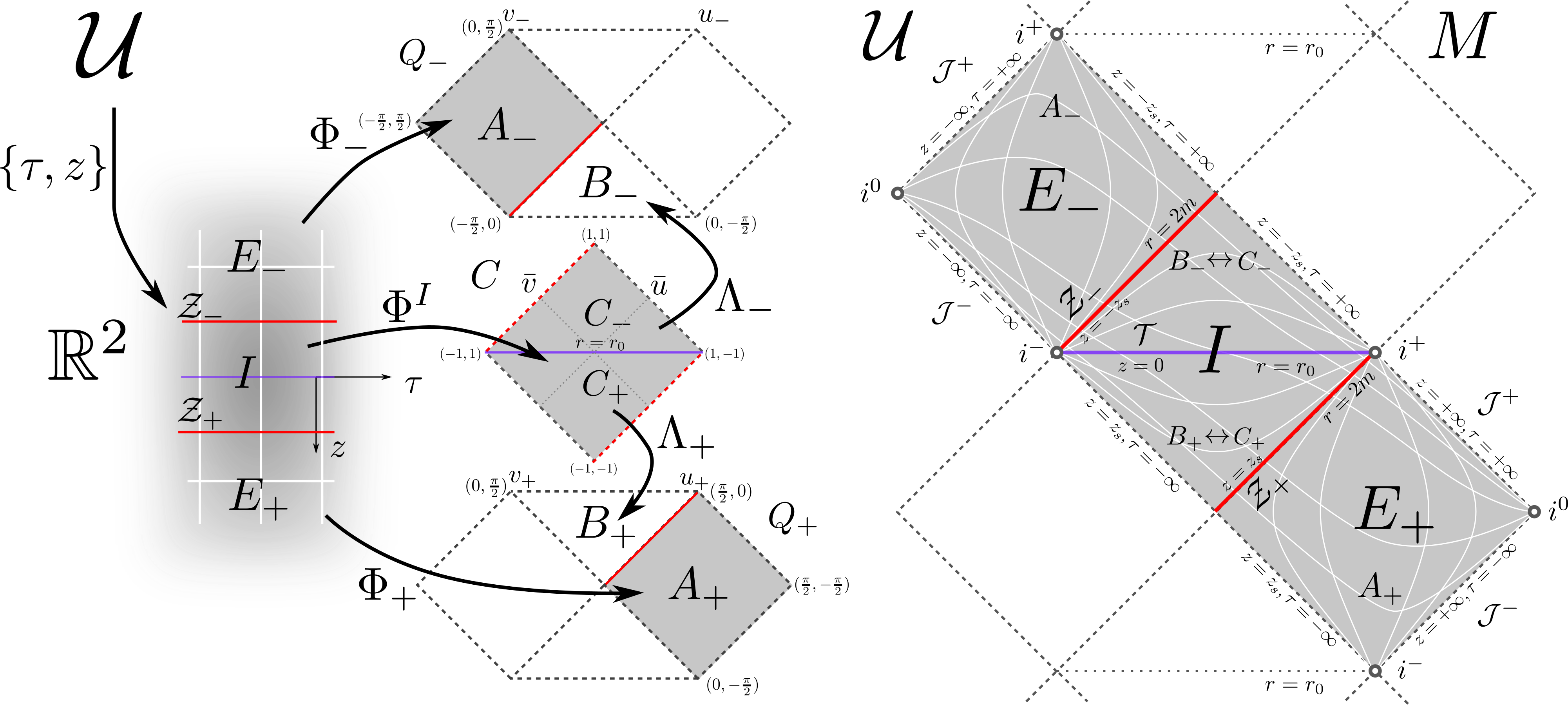}
  \caption{Penrose diagram of the domain $(\mathcal{U},g)$ (shaded) and its maximal
    analytical extension $(M,g)$ (outlined). We depict
    the diffeomorphisms that map the two exterior regions $\exterior_\sigma$
    and the interior region $\interior$, from the corresponding
    restrictions of the common chart $\{\tau,\xu\}$, to the sets $A_\sigma$ in the
    charts $\{u_\sigma,v_\sigma\}$ and $C$ in the chart $\{\uint,\vint\}$,
    respectively.}
  \label{figura1}
\end{figure*}

Here we introduce holonomy corrections
through a canonical transformation and implement a regularization
of the deformed Hamiltonian constraint. We then construct the spacetime
that solves this effective theory and obtain its global causal structure. In particular, a single chart covers a singularity-free  (black-hole/white-hole) interior region plus two asymptotically flat 
exterior regions, as depicted in Fig.~\ref{figura1}.
The main features are listed at the end of the manuscript.

In the $3+1$ setup of a manifold $M$
  based on the level hypersurfaces of some function $\time$,
  the diffeomorphism invariance of GR is encoded in four constraints:
  the Hamiltonian constraint $\widetilde\ham$, that generates
  deformations of the hypersurfaces (as a set),
  and the diffeomorphism constraint $\diff$, which has three components and
  generates deformations within the hypersurfaces.
  Spherical symmetry allows the introduction of another function $\radi$,
  constant on the symmetry orbits.
  In this case
  the two angular components of $\diff$ trivially vanish and,
  in terms of the Ashtekar-Barbero variables, we have
\begin{align*}
\diff=&
     -({\widetilde{E}^x})'{\widetilde{K}_x} +{\widetilde{E}^\varphi} ({\widetilde{K}_\varphi})',
  \\
\widetilde{\ham}
    =& -\frac{{\widetilde{E}^\varphi}}{2\sqrt{{\widetilde{E}^x}}}\left(1+\widetilde{K}_\varphi^2\right)  -2\sqrt{{\widetilde{E}^x}}{\widetilde{K}_x}{\widetilde{K}_\varphi} + \frac{(({\widetilde{E}^x})')^2}{8 \sqrt{{\widetilde{E}^x}} {\widetilde{E}^\varphi}}  \nonumber
    \\
    & - \frac{\sqrt{{\widetilde{E}^x}}}{2(\widetilde{E}^\varphi)^2} ({\widetilde{E}^x})' ({\widetilde{E}^\varphi})' +\frac{\sqrt{{\widetilde{E}^x}}}{2{\widetilde{E}^\varphi}}({\widetilde{E}^x})'',
\end{align*}
with prime the derivative with respect to $x$,
$\widetilde{E}^x>0$ and $\widetilde{E}^\varphi$ the components of the symmetry-reduced triad,
and $\widetilde{K}_x$ and $\widetilde{K}_\varphi$ their conjugate momenta.
The symplectic structure is $\{\widetilde{K}_i(\radi_a),\widetilde{E}^j(\radi_b)\}=\delta_i^j\delta(\radi_a\!-\!\radi_b)$ for $i,j\!=\!x,\varphi$.
These constraints satisfy the Poisson algebra
\begin{align*}
 \{D[f_1], D[f_2] \} &= D[f_1 f_2'-f_1'f_2],\\
  \{D[f_1], \widetilde{H}[f_2] \} &= \widetilde{H}[f_1 f_2'],\\
   \{\widetilde{H}[f_1], \widetilde{H}[f_2] \} &= D[\widetilde{E}^x(\widetilde{E}^\varphi)^{-2}(f_1 f_2'-f_1'f_2)],
\end{align*}
with $\widetilde{H}[f]:=\int\! f\widetilde{\ham}{d\radi}$ and $D[f]:=\int\!f\diff{d\radi}$.
The combination $\widetilde{H}[N]+D[\shift]$,
with Lagrange multipliers $\lapse$ and $\shift$, is the GR
Hamiltonian for vacuum in spherical symmetry,
which we will refer to as the \textit{classical} theory in the remainder.

In loop quantum gravity only the holonomies of the connection, and not the connection itself, have a well-defined operator counterpart. Hence, effective descriptions usually perform a polymerization procedure which, essentially, replaces each $\widetilde{K}_\varphi$
with a periodic function such as $\sin(\lambda \widetilde{K}_\varphi)/\lambda$. The parameter $\lambda$ encodes the discretization
of the quantum spacetime.
Nonetheless, this simple polymerization may give rise to anomalies since the deformed
constraint algebra does not generically close. Although a careful choice of the functions allows to define an anomaly-free polymerized Hamiltonian in vacuum, the presence of matter with local degrees of freedom rules out that possibility \cite{Bojowald:2015zha,Bojowald:2020unm,Alonso-Bardaji:2020rxb}.

In view of the above, the idea introduced in \cite{cantransf,Alonso-Bardaji:2021tvy} is
to consider not just modifications of
$\widetilde{K}_\varphi$ but also of its conjugate variable $\widetilde{E}^\varphi$.
For instance, if one performs the canonical transformation 
$\widetilde{K}_\varphi=\sin(\lambda {\kang})/\lambda$, 
$\widetilde{E}^\varphi={\ephi}/\cos(\lambda {\kang})$,
$\widetilde{K}_x=K_x$, and $\widetilde{E}^x=\erad$, the theory remains free of anomalies even when adding matter fields. Note that this transformation leaves invariant the diffeomorphism constraint ${\cal D}= -{\erad}'{\krad} +{\ephi} \kang'$. 
As long as $\cos(\lambda {\kang})$ does not vanish,
the canonical transformation is bijective and, essentially,
the dynamical content of the theory is the same as that given by GR.
However, the surfaces (see below)
$\cos(\lambda {\kang})=0$ may contain novel physics.
Since the Hamiltonian constraint diverges there, we regularize it,
and define the linear combination
\begin{align}
{\cal H}:=\left(\widetilde{\ham}
 +\lambda\, \sin(\lambda K_\varphi) \frac{\sqrt{\erad}\erad{}'}{2(\ephi)^2}{\cal D}\right)
  \frac{\cos(\lambda \kang)}{\sqrt{1+\lambda^2}},
   \label{H_normal}
\end{align}
along with $H[f]\!:=\!\int\!\! f {\cal H} d\radi$, so that
the canonical algebra
\begin{align}
\{D[f_1], D[f_2] \} &= D[f_1f_2'-f_1' f_2],\nonumber\\
  \{D[f_1], {H}[f_2] \} &= {H}[f_1f'_2],\label{algebra}\\
   \{{H}[f_1], {H}[f_2] \} &= D[F(f_1f_2'-f_1' f_2)],\nonumber
\end{align}
follows, with the non-negative structure function
\begin{align*}
F:=  
    \frac{\cos^2(\lambda {\kang})}{1\!+\!\lambda^2}\left(1+\Big(\frac{\lambda {\erad}'}{2{{\ephi}}}\Big)^{\!2}\right)\frac{\erad}{(E^{\varphi})^2}.
\end{align*}
Now, let us define
\begin{align*}
m:=\frac{{\sqrt{\erad}}}{2}\left(\!1+\frac{\sin^2(\lambda {{\kang}})}{\lambda^2}-\left(\!\frac{{\erad}'}{2{{\ephi}}}\!\right)^{\!\!2}\!\cos^2(\lambda {{\kang}})\!\right) ,
\end{align*}
which is a constant of motion. It is important to note now that
   the condition $\cos(\lambda {\kang})=0$ holds
if and only if  $\sqrt{E^x}=2m\lambda^2/(1+\lambda^2)$,
which is a gauge-independent
statement because $E^x$ is a scalar.
Therefore, although $\kang$ is not a scalar quantity,
$\cos(\lambda {\kang})=0$ covariantly defines surfaces on $M$.
For convenience, we introduce
$r_0:= 2m\lambda^2/(1+\lambda^{2})$, so that
\begin{align*}
   F=\left(1-\frac{r_0}{\sqrt{{\erad}}}\right)\frac{\erad}{(\ephi)^2}.
\end{align*}
From now on we will assume $m\!>\!0$ and $\lambda\!\neq\!0$, and thus
$0\!<\! r_0\!<\!2m$. The classical theory is recovered in the limit $\lambda\rightarrow0$, which implies $r_0\to0$.
Let us stress that the characteristic scale $r_0^2$ arises naturally from the constraint algebra and will show up in the model as a minimal area.

To construct a consistent geometric description,
we use the functions $\time$ and $\radi$ on $M$,
  plus the unit sphere metric $\Sunit$,
  to produce a chart $\{\time,\radi\}$ (we omit the angular part)
  in which a spherically symmetric metric $g$
  is given in the general form
\begin{align}
 ds^2=-L^2 d\time^2+q_{xx}(d\radi+ S d\time)^2+q_{\varphi\varphi}d\Omega^2.
\label{g_spher}
\end{align}
The lapse $L$, shift $S$, $q_{xx}$ and $q_{\varphi\varphi}$ depend on $\time$ and $\radi$. The unit normal
to the hypersurfaces of constant $\time$ is given by $n=L^{-1}(-\partial_{\time}+S\partial_{\radi})$.
Our purpose is to define these functions
in terms of phase-space variables in such a way that infinitesimal
coordinate transformations coincide with gauge variations.
We start by imposing that the Hamiltonian construction is based indeed on
$\time$ and $\radi$, that is,
  the Lagrange multipliers correspond to the lapse and shift,
  hence $L=\lapse$ and $S=\shift$ as functions on $M$.
Now, on the one hand, an infinitesimal change of coordinates $(\time+\xi^{\time},\radi+\xi^{\radi})$
is given by the Lie derivative of
the metric along the vector $\xi=\xi^{\time} \partial_{\time}+\xi^{\radi} \partial_{\radi}$.
On the other hand, a gauge transformation
of a function $G$ on the phase space is given by
$\delta_\epsilon G=\{G,H[\epsilon^0]+D[\epsilon^x]\}$, with gauge parameters $\epsilon^0$
and $\epsilon^x$.
Since $H$ and $D$ satisfy the canonical algebra \eqref{algebra},
these two deformations should coincide if the gauge parameters correspond
to the components of the normal decomposition of the vector $\xi$ \cite{Teitelboim73}, that is,
$\xi=\epsilon^0 n+\epsilon^x \partial_{\radi}$,
which implies the relations
$\epsilon^0=\lapse\xi^{\time}$ and $\epsilon^x=\xi^{\radi}+\xi^{\time} \shift$.
In particular, the modification of the Lagrange multiplier
$\shift$ under a gauge transformation
is given by \cite{Pons:1996av,Bojowald:2018xxu}
\begin{equation*}
\delta_\epsilon \shift= \dot{\epsilon^x}+\epsilon^x \shift^{\prime}-\shift \epsilon^{x\prime}
  -F(\lapse\epsilon^{0\prime}-\epsilon^0 \lapse'),
\end{equation*}
whereas, under infinitesimal coordinate transformations,
the shift changes as
\begin{equation*}
\delta_\xi \shift\!\!=\!\dot{\xi^{\radi}}\!+\!\dot{\shift}\xi^{\time}\!+\! \shift^{\prime}\xi^{\radi}\!+\! \shift(\dot{\xi^{\time}}\!-\!\xi^{\radi\prime})\!-\!\left[\!\frac{\lapse^2}{q_{xx}}\!+\!(\shift)^2\!\right] \!\xi^{\time\prime},
\end{equation*}
the dot being the time derivative.
The equivalence of the two variations needs $q_{xx}=1/F$,
which can be consistently imposed since
$\delta_\xi q_{xx}=\delta_\epsilon (1/F)$.
Also, we have for the lapse $\delta_\xi N=\delta_\epsilon N$.
Finally, we demand $q_{\varphi\varphi}$ to retain its classical
form, $q_{\varphi\varphi}=\erad$, which has the correct transformation properties.
The explicit details of the equivalence of gauge variations in phase space
and coordinate transformations of this construction are shown in \cite{in_preparation}.

We thus end up with the metric, c.f. \eqref{g_spher},
\begin{align}\label{effectivemetric}
{ds}^2 =-\lapse^2{d\time}^2 +
  \frac{({\ephi})^2}{{\erad}}\frac{\big({d\radi}+{\shift}{d\time}\big)^2}{1-r_0/\sqrt{\erad}}    
  +{\erad}{d\Omega}^2.
\end{align}
Compared to its classical form, it contains the term $(1-r_0/\sqrt{{\erad}})$.
Also, the precise form of ${\erad}$, ${\ephi}$, $\lapse$ and $\shift$
as functions of the coordinates will not be generically the same as in GR,
since they must solve the deformed system of
equations
$\dot{E^i}=\{E^i,H[\lapse]+D[\shift]\}$, $\dot{K_i}=\{K_i,H[\lapse]+D[\shift]\}$,
for $i=x,\varphi$, along with ${\cal H}=0$ and ${\cal D}=0$
\cite{in_preparation}.
Since our construction is consistent, different gauge choices will simply lead to different coordinate
charts (with different domains of $M$ in general) and corresponding expressions for the same metric.
Next we find the solution to that system and obtain the corresponding \textit{unique} geometry
for four different gauges.

\noindent
\textit{(a) A static region:
} Using the labels $\{\time,\radi\}=\{\tilde t,\rext\}$
for this chart, and setting $\erad=\rext^2$ and $\kang=0$  we get
\begin{equation}\label{deformedschwarzsmetric}
ds^2\!=\!-\!\bigg(\!1-\frac{{2m}}{\rext}\!\bigg){d\tilde t}^2 +\bigg(\!1-\frac{r_0}{\rext}\!\bigg)^{\!\!-1}\!\bigg(\!1-\frac{{2m}}{\rext}\!\bigg)^{\!\!-1}\!\!\!{d\rext}^2 \!+\rext^2d\Omega^2,
\end{equation}
with $\rext\in(2m,\infty)$. This region is asymptotically flat, and will describe one exterior domain.

\noindent
\textit{(b) A homogeneous region:
} We name $\{\time,\radi\}=\{T,Y\}$
and demand $\erad'=\ephi'=0$. Taking $\erad=T^2$ we obtain
\begin{equation}\label{metschwarzshom}
ds^2\!=\! -\!\bigg(\!1-\frac{r_0}{T}\!\bigg)^{\!\!\!-1}\!\!\bigg(\!\frac{{2m}}{T}-1\!\bigg)^{\!\!\!-1}\!\!\!\!{dT}^2+\bigg(\!\frac{{2m}}{T}-1\!\bigg)\!{dY}^2+T^2\! d\Omega^2\!,
\end{equation}
with $T\in (r_0, 2m)$, that will describe half of a homogeneous
Kantowski-Sachs type interior.

None of these two coordinate systems crosses the horizon at $r=2m$,
nor the instant $T=r_0$, and their domains on $M$
do not intersect. The next gauge \textit{(c)} produces a chart on a domain
that will cover two  regions \textit{(b)}, providing
  the full interior homogeneous region
  including the hypersurface $T=r_0$;
 whereas the gauge \textit{(d)} yields a chart on a domain
$\mathcal{U}\subset M$
that covers all the above.

\noindent

\textit{(c) The whole homogeneous region:
} We set $\{\time,\radi\}=\{\timeI,\xuI\}$
and demand $\erad'=\ephi'=0$ as in \textit{(b)},
but now we take $\kang=\timeI/\lambda$. Naming $\sqrt{\erad}=:\rint$, we obtain
\begin{equation}\label{g_I_full}
  ds^2=-\frac{2\rint(\timeI)^4}{m r_0} d\timeI^2\!
  +\!\left(\frac{2m}{\rint(\timeI)}-1\right)
  d\xuI^2+\rint(\timeI)^2 d\Omega^2,
\end{equation}
where $\rint(\timeI)=2mr_0/(2m\sin^2\timeI+r_0\cos^2\timeI)$
, so that
\begin{equation}
  \frac{2m}{\rint(\timeI)}-1=\left(\!\frac{2m}{r_0}-1\!\right)\sin^2\timeI,
  \label{r_T}
\end{equation}
and the range of coordinates is restricted to
$\timeI\in (0, \pi)$. This region will describe the full homogeneous
Kantowski-Sachs type interior, and contains the
spacelike hypersurface $\rint=r_0$, located at $\timeI=\pi/2$.


\noindent
\textit{(d) The covering domain $\mathcal{U}$:}
Now $\{\time,\radi\}=\{\tau,\xu\}$, and
we impose $\dot{\erad}=0$, $(\erad')^2=4\erad(1-r_0/\sqrt{\erad})$ and $\ephi=\erad'/2$.
Renaming for simplicity $\sqrt{\erad}=:r$,
\begin{equation}
  ds^2\!=\!-\bigg(\!1-\frac{{2m}}{r(\xu)}\!\bigg) {d\tau}^2  
  +2\sqrt{\frac{\!{2m}}{r(\xu)}}{d\tau}{d\xu} + {{d\xu}^2} +r(\xu)^2 d\Omega^2\!,
  \label{metric:tau_x}
\end{equation}
with $(\tau,\xu)\in\mathbb{R}^2$. The function $r$ in this chart,
$r(\xu)$, is even $r(-\xu)=r(\xu)$ and it is implicitly given by
\[
|\xu|= r(\xu)\sqrt{1-\frac{r_0}{r(\xu)}}+r_0\log{\Bigg(\sqrt{\frac{r(\xu)}{r_0}}+\sqrt{\frac{r(\xu)}{r_0}-1}\Bigg)}.
\]
Observe that $r(0)=r_0>0$ is its only minimum and
$r(\xu)$ is analytic on $\mathbb{R}$, with image on $[r_0,\infty)$.

The chart $\{\tau,\xu\}$ thus maps some
domain $\mathcal{U}\!\subset\! M$ 
to the whole plane $\mathbb{R}^2$.
In the search 
for the global structure of $(\mathcal{U},g)$
we will produce appropriate coordinate transformations
so that \eqref{metric:tau_x} takes the explicit
conformally flat form on the $(\tau, z)$-plane, see \eqref{g:uv_ext} and \eqref{g:uv_int},
that will coincide with \eqref{deformedschwarzsmetric}, \eqref{metschwarzshom}, and \eqref{g_I_full} on their corresponding domains.
This will show that $(\mathcal{U},g)$ covers any such
static region \textit{(a)} and
homogeneous  regions \textit{(b)} and \textit{(c)}.
The procedure will end by proving
that $(\mathcal{U},g)$ contains exactly one
globally hyperbolic interior domain composed of
one homogeneous region \textit{(c)}, which
  covers two  regions \textit{(b)},
and two exterior regions \textit{(a)}.
This whole process, along with the resulting
spacetime diagram, is sketched in Fig.~\ref{figura1} (for further details see \cite{in_preparation}).

We first define the sets:
$\exterior:=\{r>2m\}\cap \mathcal{U}$,
$\interior:=\{r<2m\}\cap \mathcal{U}$,
$\horizon:=\{r=2m\}\cap\mathcal{U}$, and
$\mathcal{T}:=\{r=r_0\}\cap\mathcal{U}\subset \interior$.
Then we use the chart $\{\tau,\xu\}$ to decompose
these sets (except $\mathcal{T}$) by taking their restrictions
under the sign function $\sgn(z)$,
and use the notation
$D_\sigma:= D|_{\sgn(\xu)=\sigma}$ (with $\sigma=\pm 1$) for any domain $D$.
In particular, $\exterior=\exterior_-\cup \exterior_+$  is
disconnected
and $\interior=\interior_-\cup\mathcal{T}\cup \interior_+$
is a connected set.
To ease the notation we will use the same letter for
the domain in $\mathcal{U}$ and its image on $\mathbb{R}^2$
under the chart $\{\tau,\xu\}$. For instance,
$\exterior_+$ also stands for the half plane
$\xu\in(\xu_s,\infty)$ in $\mathbb{R}^2$, where $\xu_s$ is the positive root
of $r(\xu_s)=2m$, and $\interior$ is also the stripe $\xu\in(-\xu_s,\xu_s)$.

With the auxiliary
$\alpha:=4m(1-\frac{r_0}{2m})^{-1/2}$, $\varepsilon\!=\!\pm 1$ and
\begin{align*}
 &R_\varepsilon(r):=\alpha
      \log\left(\sqrt{r_0}\frac{\left|\sqrt{r-r_0}+\varepsilon\sqrt{2m-r_0}\right|}
        {\sqrt{2m}\sqrt{r-r_0}+\sqrt{r}\sqrt{2m-r_0}}\right)\\
&+\left(\sqrt{r}-\varepsilon \sqrt{8m}\right)\!\sqrt{r-r_0}+\!(4m \!+\! r_0)\!\log\!\left[\!\sqrt{\frac{r}{r_0}}\!+\!\sqrt{\frac{r}{r_0}-1}\right]
  \end{align*}
we construct
$R_U(r):=R_+(r)$, $R^\exterior_V(r):=R_-(r)|_{r>2m}$ and
$R^\interior_V(r):=R_-(r)|_{r<2m}$.
Since $R_U(r_0)=0$, it is easy to check that
$U(\tau,\xu):=\tau+\sgn(\xu)R_U(r(\xu))$
is analytic on the whole plane.

Let us first work out the causal structure of the exterior regions
$\exterior_\sigma$. On each $\exterior_\sigma$
we define the respective function
$ V_\sigma(\tau,\xu)=\tau-\sigma R^\exterior_V(r(\xu))$,
analytic on its domain,
and use $U_\sigma:=U|_{\sgn(\xu)=\sigma}$ to construct the diffeomorphisms
$\Phi_\sigma:\{\tau,\xu\}|_{\exterior_\sigma}\!\to\!\{u_\sigma,v_\sigma\}$ by
\begin{align*}
 &u_\sigma=\sigma\arctan\exp\left[\frac{\sigma}{\alpha}U_\sigma(\tau,\xu)\right],\\
 &v_\sigma=-\sigma\arctan\exp\left[-\frac{\sigma}{\alpha}V_\sigma(\tau,\xu)\right],
\end{align*}
that map, respectively, the domains $\exterior_+$ and $\exterior_-$
to the regions $A_+=\{u_+\in(0,\pi/2), v_+\in(-\pi/2,0)\}$ and
$A_-=\{u_-\in(-\pi/2,0), v_-\in(0,\pi/2)\}$.
In the charts $\{u_\sigma,v_\sigma\}$
the metric, c.f. \eqref{metric:tau_x}, reads
\begin{align}
  ds^2=\frac{\ff(r(u_\sigma,v_\sigma))}{\cos^2u_\sigma\cos^2v_\sigma }du_\sigma dv_\sigma+r(u_\sigma,v_\sigma)^2\Sunit,
  \label{g:uv_ext}
\end{align}
with
\begin{align}
  \!\!&\ff(r)\!:= \!-\frac{2m\alpha^2}{r}\!\left(\!\!\sqrt{1\!-\!\frac{r_0}{r}}
      \!+\!\sqrt{1\!-\!\frac{r_0}{2m}}\,\right)^{\!\!2}\!\!
\exp\!\left[-\frac{2r}{\alpha}\sqrt{1\!-\!\frac{r_0}{r}}\,\right]\quad\mbox{}\nonumber \\
    &\times \!\!
\left(\!1\!+\!\sqrt{1\!-\!\frac{r_0}{r}}\right)^{\!\!-\!\sqrt{1-\frac{r_0}{2m}}(2+\frac{r_0}{2m})}\!\!\!\left(\frac{r_0}{r}\right)^{\!\!\sqrt{1-\frac{r_0}{2m}}(1+\frac{r_0}{4m})-1}\!\!,\!\!\label{Gamma}
\end{align}
and $r(u_\sigma,v_\sigma)$ satisfies
\begin{equation}
  \label{eq:r_uv}
  \tan u_\sigma \tan v_\sigma=\left(1-\frac{2m}{r}\right)\frac{\alpha^2}{\ff(r)}
  =:\Upsilon(r).
\end{equation}
$\ff(r)$, as defined, is finite and negative
  on $r\in[r_0,\infty)$
  and satisfies $\ff(r)\Upsilon'(r)=2\alpha(1-r_0/r)^{-1/2}$.
  Hence, 
$\Upsilon$ is a strictly decreasing function of $r$
  with $\Upsilon(r_0)=1$ and $\Upsilon(2m)=0$.
  For each $\exterior_\sigma$ the set $A_\sigma$ thus provides the usual Penrose
diagram for the Schwarzschild exterior.
Moreover, on each $\exterior_\sigma$ the change
$\{\tau,\xu\}|_{\exterior_\sigma}\!\to\!\{\tilde t,\rext\}$, given by $\rext=r(\xu)$ and 
$
\tilde t=\tau+\frac{\sigma}{2}\left(R_U(r(\xu))-R^\exterior_V(r(\xu))\right)$,
produces a chart in which the metric, c.f. \eqref{metric:tau_x},
reads as \eqref{deformedschwarzsmetric}.
This shows $\mathcal{U}$ covers two exterior regions isometric to \textit{(a)}.

We proceed similarly for the interior region $\interior$.
First, we define $ V^\interior(\tau,\xu):=-\tau+\sgn(\xu) R^\interior_V(r(\xu))$,
which is analytic in $\xu\in(-\xu_s,\xu_s)$
(note that $R^\interior_V(r_0)=0$), and $U^I:=U|_\interior$.
 It can be checked that $R_U(r)+R^\interior_V(r)<0$ for 
   $r\in(r_0,2m)$, and therefore $\sgn(U^I+V^I)=-\sgn(\xu)$.
The diffeomorphism
$\Phi^I:\{\tau,\xu\}|_{\interior}\to\{\uint,\vint\}$
\[
  \uint=\tanh\left[\frac{1}{2\alpha}U^\interior(\tau,\xu)\right],\quad
  \vint=\tanh\left[\frac{1}{2\alpha}V^\interior(\tau,\xu)\right],
\]
maps $\interior$
to $C=\{\uint\in(-1,1), \vint\in(-1,1)\}$.
In this chart the metric, c.f. \eqref{metric:tau_x}, reads
\begin{equation}
  \!\!\!ds^2\!=\!{\left(\!1\!-\!\frac{2m}{r(\uint,\vint)}\!\right)\!\frac{\alpha^2}{(1\!-\!\uint^2)(1\!-\!\vint^2)}}d\uint d\vint
  +r(\uint,\vint)^2\Sunit\!,
  \label{g:uv_int}
\end{equation}
where $r(\uint,\vint)$ satisfies
\begin{align*}
  -\left|\frac{\uint+\vint}{1+\uint\vint}\right|
    =\tanh\left[\frac{1}{2\alpha}(R_U(r)+R^\interior_V(r))\right].
\end{align*}
Since  $R_U(r_0)=R^\interior_V(r_0)=0$
the curve $r=r_0$ is mapped to the horizontal
line $\uint+\vint=0$.
Further, 
  we have
  $\sgn(U^I+V^I)=\sgn(\uint+\vint)$ and hence
  $\sgn(\uint+\vint)=-\sgn(\xu)$.
  Therefore, each set of constant $r\in(r_0,2m)$ corresponds to
two curves of constant $(\uint+\vint)/(1+\uint\vint)$
that go from $(\uint,\vint)=(-1,1)$ to $(\uint,\vint)=(1,-1)$
through positive (negative) values of $\uint+\vint$ for $\sgn(\xu)=-1$
($\sgn(\xu)=1$). For finite $\tau$, as $r\to 2m$,
the function $R_U$ remains bounded whereas $R^\interior_V\to -\infty$.
Hence points approaching $\horizon_\sigma$ from $\interior$
attain $\vint\to -\sgn(\xu)$, while $\uint$ runs over
its whole range. The set $C$ provides the Penrose diagram for $\interior$.
The change $\{\tau,\xu\}|_{\interior}\to \{\timeI,\xuI\}$ defined
  by $\tau=\xuI-\frac{\sgn(z)}{2}\left(R_U(\rint(\timeI))-R^\interior_V(\rint(\timeI))\right)$
  and $\rint(\timeI)=r(\xu)$, that imply $\sgn(\xu)=\sgn(\cos\timeI)$ and is
  explicitly given by
      \begin{align*}
        \tau=&\xuI-\alpha\,\mbox{artanh}\left[\sqrt{\frac{\rint(\timeI)}{2m}}\cos\timeI\right]
               +\frac{16 m^2}{\alpha}\sqrt{\frac{\rint(\timeI)}{2m}}\cos\timeI,\\
        \xu=&r_0\,\mbox{artanh}\left[\frac{4m}{\alpha}\cos \timeI\right]
              +\frac{\alpha}{8m^2}\rint(\timeI)\cos\timeI
      \end{align*}
      with $\eqref{r_T}$,
      is one-to-one and
      takes \eqref{metric:tau_x} to the form \eqref{g_I_full}.
      This shows that $\interior\subset\mathcal{U}$ is isometric
      to the region \textit{(c)}. Observe that $z=z_s$, i.e. $\horizon_+$, is recovered
      for $\timeI=0$; $z=-z_s$, i.e. $\horizon_-$, for $\timeI=\pi$; and
      $z=0$, that is $\mathcal{T}$, for $\timeI=\pi/2$.
    Further, on each $\interior_\sigma$, the change
    $\{\tau,\xu\}|_{\interior_\sigma}\to \{T,Y\}$, given by $T=r(\xu)$ and
    $Y=\tau+\frac{\sigma}{2}\left(R_U(r(\xu))-R^\interior_V(r(\xu))\right)$,
    renders the metric, c.f. \eqref{metric:tau_x}, in the form \eqref{metschwarzshom}. Therefore
      $\interior$, and thus also the region \textit{(c)}, cover
      two regions \textit{(b)}.

Finally, we use that $\Upsilon(r)$
strictly decreases on $[r_0,\infty)$
with maximum $\Upsilon(r_0)=1$, ensuring
\eqref{eq:r_uv} has a solution for $r$ everywhere on  $\tan u_\sigma \tan v_\sigma\leq 1$.
Therefore
each set $A_\sigma$ can be extended to the Kruskal-Szekeres-type regions
$Q_\sigma$ (see Fig.~\ref{figura1}).
The purpose of these extensions is twofold. Firstly,
the sets  $C_\sigma$ can be mapped respectively to
$B_+:=\{u_+,v_+\in\!(0,\pi/2);u_+\!+\!v_+<\pi/2\}\subset Q_+$ and
$B_-:=\{u_-,v_-\in\!(-\pi/2,0);u_-\!+\!v_-\!>\!-\pi/2\}\subset Q_-$
with the diffeomorphisms
 $\Lambda_\sigma:C_\sigma\to B_\sigma$
\[
  u_\sigma=\sigma\arctan\left(\frac{1+\uint_\sigma}{1-\uint_\sigma}\right)^{\sigma},
  v_\sigma=\sigma\arctan\left(\frac{1+\vint_\sigma}{1-\vint_\sigma}\right)^{\sigma},
\]
in order to define the extended charts $\{u_\sigma,v_\sigma\}_{Q_\sigma}$
so that they map all $p\in \interior_\sigma$
by $\Lambda_\sigma\circ\Phi^\interior(p)$ to the
respective point on $B_\sigma$. This ends the contruction of the full Penrose
diagram for $(\mathcal{U},g)$.
Secondly, we extend $(\mathcal{U},g)$
to two Kruskal-Szekeres-type analytic regions,
$Q_+$ and $Q_-$, by adding their remaining halfs.
 These can be used to build up the maximal analytic extension of $M$
  in the usual periodic fashion, and show
  that $M$ is geodesically complete \cite{in_preparation}.

  All test particles that cross the horizon at $r=2m$, arrive
  to $r_0$, continue towards negative values of $\xu$
  with increasing values of $r$, and cross again $r=2m$ after a finite proper time. In particular, radial infalling particles at rest at infinity take a time
  $\frac{8}{3}(m+r_0)(1-\frac{r_0}{2m})^{1/2}$
  to cross the interior region.
The singularity in Schwarzschild, at $r=r_0=0$, is not present here
and the curvature is bounded. In particular, the
curvature scalars take their maximum value at $r=r_0$.
For instance, the Ricci scalar $R= 3mr_0 /r^4$ is everywhere positive.
Note that even if quantum-gravity effects (parametrized
  by $r_0$) are present outside the horizon, they decay as one moves
  to low-curvature regions.

  The computation of the expansions of ingoing
  and outgoing radial null congruences shows, as expected,
  that the spheres of constant $\time$ and $\radi$
  are non-trapped in the exterior region $r>2m$,
  and that $r=2m$ is indeed a horizon.
 Moreover, in the interior region $r_0<r<2m$
both expansions have the same sign, given by $-\sgn(z)$,
and vanish at $r=r_0$. Therefore, in $\interior_+$ ($\interior_-$)
those spheres are trapped (anti-trapped)
while in $\mathcal{T}$, they have zero mean curvature.
  In fact, the hypersurface $r=r_0$ itself is minimal,
  reflecting the mirror symmetry $\xu\to -\xu$.

  Therefore, as one expects for a singularity resolution,
  some of the eigenvalues of the Einstein tensor ${G^a}_b$
  must attain negative values on $\interior$.
  Indeed, if one interprets $G^a{}_b$ as an effective energy-momentum
  tensor, the eigenvalues on the angular part
  would define an angular pressure $(r-m)r_0/(2r^4)$.
  On $\exterior$ one would get a positive energy density $2m r_0/r^4$
  and a negative radial pressure $-r_0/r^3$,
  while on $\interior$ the energy density would be $r_0/r^3$ and the radial pressure $-2m r_0/r^4$.
  With these values it is easy to check that none of the \textit{geometric}
  energy conditions are satisfied at any point except at the horizon.
 However, let us recall that $(M,g)$ solves the vacuum equations,
  thus satisfying trivially all the \textit{physical} energy conditions.

Let us finally summarize the main features of this effective quantum
black-hole model:
$(i)$ The brackets between deformed constraints
vanish on-shell and thus form an anomaly-free algebra.
$(ii)$ We provide a consistent, hence covariant, geometric setup
so we can talk of a \textit{metric that solves the system}.
  Different gauge choices on the phase space simply provide
  different charts (and domains) of some spacetime $(M,g)$,
  with corresponding expressions
  for the same metric tensor.
$(iii)$ A convenient choice of gauge provides a single chart that
  covers a domain $(\mathcal{U},g)$ with global structure shown in Fig.~\ref{figura1},
which represents a globally hyperbolic interior (black-hole/white-hole) region and two asymptotically flat exteriors of equal mass.
$(iv)$ We have produced the maximal analytical extension $(M,g)$.
$(v)$ Quantum-gravity effects introduce a length scale $r_0>0$, that defines
a minimum of the area of the orbits of the spherical symmetry,
and removes the classical singularity.
More precisely, the surface $r=r_0$ is just a minimal hypersurface between a trapped
and anti-trapped region, and all causal geodesics cross it in finite time.
$(vi)$ All curvature scalars are bounded everywhere.
$(vii)$ Quantum-gravity effects die off as we move to
low-curvature regions.
$(viii)$ Schwarzschild is recovered for $r_0=0$ and Minkowki for $m=0$.

\noindent{\bf Acknowledgments}
We acknowledge financial support from the Basque Government Grant \mbox{No.~IT956-16}
and from the Grant FIS2017-85076-P, funded by MCIN/AEI/10.13039/ 501100011033 and by “ERDF A way of making Europe”.
AAB is funded by the FPI fellowship \mbox{PRE2018-086516} of the Spanish MCIN.

\bibliography{refs}

\end{document}